\documentclass[draft,grl]{AGUTeX}

\authorrunninghead{ESELEVICH AND ESELEVICH}

\titlerunninghead{A FORMATION MECHANISM FOR THE TYPE II RADIO EMISSION}

\authoraddr{V. G. Eselevich,
Institute of Solar-Terrestrial physics, Lermontova str. 126a,
P.O.Box 291, Irkutsk, 664033, Russia.}

\authoraddr{M. V. Eselevich,
Institute of Solar-Terrestrial physics, Lermontova str. 126a,
P.O.Box 291, Irkutsk, 664033, Russia.}

\begin{document}

%
%

\title{A formation mechanism for the type II radio emission in the solar corona unrelated to shock waves}
%

%
%



\authors{V. G. Eselevich \altaffilmark{1}
and M. V. Eselevich \altaffilmark{1}}

\altaffiltext{1}{Institute of Solar-Terrestrial physics, Irkutsk, Russia}






%
%


\begin{abstract}
Mark 4, COR1/STEREO and LASCO/SOHO data analysis shows that at least a portion of type II radio bursts
observed in the corona occurs in the presence of a CME, but in the absence of a shock ahead of them.
A drift current instability in the CME frontal structure is discussed as a possible cause of such bursts.
\end{abstract}

%
%

%

\begin{article}

%
%

\section{Introduction}
Experiments have revealed the existence of a turbulence zone
(``foreshock'') ahead of both the near-Earth shock
\citep{cairns1999} and interplanetary shocks travelling in the
heliosphere \citep{bale1999}. In the foreshock, there are fluxes
of energetic particles (electrons and ions) which move away from
the front along undisturbed magnetic field. They are the most
energetic portion of the plasma heated in the shock front. As a
result of beam instability evolution, electron fluxes excite
Langmuir oscillations at the plasma electronic frequency. Due to
the Rayleigh and Raman scattering \citep{zaitsev1965}, these
oscillations are transformed, respectively, into the first and
second harmonics of the type II radio emission at the single and
double plasma electron frequencies \citep{kuncic2002}. Direct
observations of the synchronous onset of increased Langmuir
oscillations ahead of the interplanetary shock front as well as
type II radio bursts at the same frequencies \citep{bale1999}
confirm the existence of this process. It is suggested that a
similar process may take place for shocks excited by coronal mass
ejections (CMEs). No direct evidence has so far been obtained,
however, due to difficulties involved in recording a shock front
in the corona. Recently developed methods for directly recording
shock fronts in the corona \citep{eselevich2008,eselevich2010}
allow one to come closer towards solving this problem.

This study aims to demonstrate that, at least, a portion of type
II radio bursts observed in the corona occur in the presence of
CMEs, but in the absence of a shock ahead of them. A possible
mechanism is discussed for the origin of such an emission
associated with a drift current instability evolving in the
leading edge of the CME frontal structure.

\section{Input data and their representation}
Our analysis employed coronal images obtained by Mark 4 (Mauna Loa Solar Observatory, http://mlso.hao.ucar.edu),
COR1 (STEREO) \citep{howard2008} and LASCO C2 (SOHO) coronagraphs \citep{brueckner1995}. The data were represented in the form of difference brightness (polarization or full) $\Delta P(t) = P(t) - P(t_0)$, where $P(t)$ is the coronal brightness at moment $t$ corresponding to the event under consideration, and $P(t_0)$ is the undisturbed brightness at fixed moment $t_0$, selected as being well before the event start.

Difference brightness images were used to inspect the dynamics of a CME and its disturbed zone. Representations in the form of isolines $\Delta P$, as well as sections both along the Sun radius at fixed position angles $PA$ and non-radial sections at various instants were employed. Position angle $PA$ is measured counterclockwise from the North Pole in the images.

\section{Data analysis method}
The analysis method is based on the results of recording the shock
front in the corona \citep{eselevich2010}. Let us briefly review
these results on the example of two CMEs with greatly different
velocities: 5 May 1997 and 20 September 1997. Both these events
are limb CMEs. Velocity with respect to the solar wind, $u$, for
these CMEs is, respectively, 150 km/s and 700 km/s ($u = V -
V_{SW}$, where $V$ is the measured CME velocity, and $V_{SW}$ is
the solar wind velocity in the streamer belt). Figures~1A and B
depict these CMEs in the form of difference brightness isolines,
at a certain instant for each event. The frontal structure is
nearly circular in shape (dashed circumferences in Figures~1A and
1B) for both the CMEs. The center of the circle, O, is $R_C$ away
from the center of the Sun along the position angle $PA_C$. The
direction from the frontal structure center is defined by angle
$\alpha$, measured counterclockwise from the CME direction
(Figure~1B).

The differences between the slow and fast CMEs are conspicuous in
both the images (Figures~1A and B) and the difference brightness
profiles $\Delta P(R)$ plotted along the T³+ propagation direction
(section 1 in Figures~1A and B). These sections in Figures~1C and
D are plotted in the frontal structure coordinate system, which
allows the formation of a shock discontinuity to be traced
practically unambiguously in both space and time [M. Eselevich,
2010]. One can see from Figure~1 that:

{\bf In the case of a slow CME ($u \approx 150$ km/sec),} the
difference brightness isolines are extend in the CME propagation
direction. The difference brightness profile $\Delta P(R)$ in the
disturbed zone ahead of the CME (black circles in Figure~1C)
smoothly decreases with distance. There is practically no
disturbed zone in the transverse direction $\alpha = -70^\circ$
(section~2, light circles).

{\bf In the case of a fast CME ($u
\approx 700$ km/sec),} the difference brightness isolines are
nearly circular in shape. A discontinuity (jump) on the scale of
$\approx 0.25$ R$_\odot$ (where R$_\odot$ is the solar radius) is
observed in the frontal part of the disturbed zone in the $\Delta
P(R)$ profile (black circles in Figure~1D). Its position is
indicated by a dashed bold line in Figure~1B.

This discontinuity was identified with the shock front based on
the analysis of more than thirty limb CMEs with velocities $V
\approx 200 \div 3000$ km/s. The analysis allowed the following
steady regularity (law) to be discovered: ``A disturbed zone
extended along the CME propagation direction exists ahead of a
coronal mass ejection when its velocity $u$ with respect to the
surrounding coronal plasma is below a certain critical velocity,
$u_C$. Shock formation ahead of the CME frontal structure in the
vicinity of CME propagation direction depends on whether local
inequality $u(R) > u_C \approx V_A(R)$ holds, which can be true at
various distances $R \geq 1.5$ R$_\odot$ from the Sun center.
Here, $V_A(R)$ is the local Alfv\'{e}n velocity of slow solar wind
in the streamer belt calculated in \cite{mann1999}''.

The local Alfv\'{e}n velocity, $V_A(R)$, is shown as a solid curve
in Figure~2, with $V_{SW}(R)$ plotted as a dotted curve. These
statements serve as a basis for identifying shocks in this study
and their relation to the presence or absence of type II radio
emission.

\section{Event selection}
The following three CMEs were selected for the analysis: 25 March
2008, 31 December 2007 and 3 June 2007. These events were
investigated in detail in \cite{gopalswamy2009} and have the
following peculiarities:
\begin{enumerate}
\item These CMEs occur close to the limb when observed by all the instruments ("limb" events). The velocity values measured by Mark4, COR1a, b and C2 differ only slightly and are close to the radial velocity.
\item The white-light CMEs have the simplest and three-part structure most convenient for investigation: frontal structure (FS), decreased brightness region (cavity) and bright core \citep{illing1986}.
\item Their propagation in the corona was accompanied by meter and decameter type II radio bursts.
\end{enumerate}

\section{Analysis of shock presence/absence for three CMEs associated with type II radio emission}

{\bf CME 1, 25 March 2008.} Figure~3 shows the difference polarization
brightness profiles $\Delta P(R)$ along the direction close to the CME motion
axis ($PA \approx 100^\circ$), drawn using the Mark~4 data for three
consecutive instants. Light circle profiles correspond to 18:41, directly prior
to the CME appearing within the Mark~4 field of view (undisturbed corona).

The position of the point from which it was drawn (parameters $PA_C$ and $R_C$
specifying the approximate center of the frontal structure) and the direction
(angle $\alpha$) are indicated for each profile. For convenient comparison, the
distances $R$ from the Sun center are marked on the x axis. Additionally, the
dashed bold line in the upper panel of Figure~3 depicts the profile $\Delta
P(R)$ at 18:47 in the direction $\alpha = 95^\circ$ (sideward with respect to
the CME motion direction). Its x position is calculated taking into account
that the shape of the frontal structure is close to a circumference. In this
direction, there is practically no disturbed zone ahead of the frontal
structure, the frontal structure boundary (current sheet) having the size
$\delta_I \sim 0.03$ R$_\odot$ (light gray on the profile), i.e. being close to
the Mark~4 spatial resolution $K \approx 0.02$ R$_\odot$. The brightness jump
$\delta P_I$ at the frontal structure boundary has an amplitude close to the
brightness jump amplitude $\delta P_F$ at the shock front.

In the region of small $\alpha$, there is a disturbed zone ahead of the frontal
structure (shown by crosshatching in the profiles), bounded by a shock in the
forefront (dark gray). (Shock propagation is seen in the following two plots in
Figure~3). The disturbed zone being, in this direction, behind the shock front,
it appears to include shock-heated plasma.

The shock front velocity, $u$, exceeds $V_A$ (black squares in Figure~2) at all
the distances. According to \cite{gopalswamy2009} the meter type II radio
emission is recorded when the forefront of the CME, corresponding to the shock,
is at 1.7 R$_\odot < R < 2.2$ R$_\odot$ (this part is crosshatched and labeled
1 in Figure~2). The decameter type II emission ceases to be recorded when the
shock reaches $R \approx 3.6$ R$_\odot$ (vertical straight line in Figure~2).
Hence the type II radio burst is accompanied by CME driven piston shock
propagation, in this case.

{\bf CME 2, 31 December 2007.} Figure~4A shows the difference brightness
profiles based on the COR1a data at consecutive instants for this event. Unlike
the profiles in Figure~3, they are drawn in the frontal structure coordinate
system, namely [Eselevich M, 2010]:
\begin{enumerate}
\item each profile is normalized to its maximum value in the vicinity of the frontal structure (i.e. the maximum value of the normalized profile is 1);
\item each profile (except the very earliest one) is shifted along the x axis so that the positions of the frontal structure coincide for all the profiles.
\end{enumerate}
Horizontal hatching shows the position of the frontal structure (the region of
the normalized profile values $> 0.5$) in Figure~4A. For the initial instant
01:00, the difference brightness profiles for two angles $\alpha$: $30^\circ$
(diamonds) and $50^\circ$ (dashed bold curve) –- are shown in Figure~4A. The
profiles are practically similar. No disturbed zone is perceptible ahead of the
frontal structure for both of these directions. The steepest segment of the
current sheet at the frontal structure boundary (light gray) in the direction
$\alpha = 50^\circ$ has a characteristic spatial scale $\delta_I \sim$
0.02-0.03 R$_\odot$, comparable to the COR1 spatial resolution $K \approx
0.016$ R$_\odot$. At the next instant 01:05 (asterisks in Figure~4L) the
disturbed zone increases only slightly in the direction $\alpha = 30^\circ$.
Thus the current sheet of the frontal structure remains practically undisturbed
prior to this instant. Its velocity $u$ only slightly exceeds $V_A$ (light
triangles in Figure~2). At the subsequent instants (01:10 and 01:15, not shown
in Figure~4L), the disturbed zone continues to gradually increase, a jump -- a
shock with the front width $\delta_F$ -- eventually forming in the very
forefront of the zone. It is shown in dark gray color for the instants 01:20
(crosses) and 01:30 (triangles). Such a change in the profile is absent in the
direction $\alpha = 50^\circ$.

Unlike the faster CME 1, where a horizontal segment of the $\Delta P(R)$
profile (i.e. $\Delta P(R) \approx$ const, Figure~3) is observed directly
behind the front, this event exhibits no steady state directly behind the shock
front even for the very latest instant 01:30 (triangles), since the difference
brightness increases as the distance decreases. This may possibly mean that the
shock front has been in the forming stage up to the instant in question. The
encircled solid triangles correspond to it in the plot $u(R)$ in Figure~2. It
is only at large distances $R > 4$ R$_\odot$ (non-encircled black triangles in
Figure~2) that a steady shock front is recorded (a segment with $\Delta P(r)
\approx$ const is observed directly behind the front). The meter type II radio
emission \citep{gopalswamy2009} is recorded when the forefront of  CME 2,
corresponding either to the current sheet at the frontal structure boundary or
to the forming shock wave, is at 1.5 R$_\odot < R < 2.6$ R$_\odot$
(crosshatched segment labelled 2 in Figure~2). The decameter type II emission
ceases to be recorded when the forefront of the CME, corresponding to the
shock, reaches $R \approx 3.5$ R$_\odot$ (vertical line in Figure~2).

{\bf CME 3, 3 June 2007.} The difference brightness profiles $\Delta P(r)$ for
this event are shown in Figure~4B. These profiles are based on the COR1a data,
in the frontal structure coordinate system in the direction $\alpha = 0^\circ$
for the two consecutive instants 09:35 (crosses) and 09:55 (triangles). These
two profiles exhibit the evolution of the disturbed zone (shown by
crosshatching). There is no shock. For comparison, the dashed bold line shows
the profile $\Delta P(r)$ at 09:45 plotted in the sideward direction $\alpha =
-60^\circ$. The steepest segment, at the frontal structure boundary, in this
profile has a difference brightness jump, $\delta P_I$, which is larger than
the half of the maximum brightness on the scale $\delta_I \sim 0.02$~R$_\odot$,
comparable to the COR1 spatial resolution.

The velocity of the forefront of the disturbed zone is lower than $V_A$ (light
circles in Figure~2) at practically all the distances. The meter and decameter
type II radio emission (shown by crosshatching 3 and vertical line in Figure~2)
is recorded when the forefront is at 1.7 R$_\odot < R < 2.6$ R$_\odot$, with no
shock signatures observable in the difference brightness profiles.

Therefore, even though the type II radio bursts were observed in all the three
events in question, the initial stage of the burst was not related to the shock
in the CME~2 case, while no shock was observed in the CME~3 case. Thus,
analysis of the last two CMEs -- 2 and 3 -- allows one to conclude that, at
least, a portion of type II bursts may occur in the absence of CME-excited
shocks. Can the observed radio bursts be caused by a different source?

\section{Discussion of a possible shock-unrelated mechanism for the type II radio emission in the solar corona}

Note, first of all, that a magnetic field jump $\delta B_I$ must correspond to
a brightness jump $\delta P_I$ on the same spatial scale $\delta_I$ at the CME
frontal structure boundary under the conditions of rarefied magnetized coronal
plasma. Drift current instability at the CME frontal structure boundary
(similarly to drift current instability at the laminar shock wave front
\citep{zaitsev1965}) may serve as the possible cause of plasma oscillations
(subsequently transforming into radio emission). The condition $V_d > V_{Te}$
must be satisfied for such oscillations to be excited, where $V_d$ and $V_{Te}$
are, respectively, the drift and thermal electron velocities \citep{chen1984}.
Let us estimate their values.

The drift velocity can be estimated from the Maxwell equation: $V_d \approx
\delta B_Ic/4\pi eN\delta_I$, where $\delta B_I$ and $\delta_I$ are,
respectively, the amplitude and scale of the magnetic field jump, $c$ is the
light speed, $N$ is the electron density, $e$ is the electron charge. If the
CME is regarded as a ``magnetic barrier'' moving in the solar wind, then
$\delta_I \sim u/\omega_{ei}$ in rarified plasma, where $\omega_{ei} = e\delta
B_I/c(m_im_e)^{0.5}$ is the hybrid cyclotron frequency \citep{longmire1963}.
Here, $c$ is the light speed, $e$ is the electron charge, $m_i$ and $m_e$ are
the ion and electron mass, respectively. Assuming $u \sim V_A$, we obtain
$\delta _I \sim c/\omega_{pe}$, where $\omega_{pe} = (4\pi Ne^2/m_e)^{0.5}$ is
plasma electron frequency. This allows the electron drift velocity in the
current sheet at the CME frontal structure boundary to be estimated as:

 \[ V_d \approx \delta B_Ic/4\pi eN\delta_I \approx \delta B_I/(4\pi Nm_e)^{0.5}. \]

Let $B_0$ be undisturbed magnetic field directly ahead of a CME. According to
the calculations by \cite{chen1996} and heliospheric observations by
\cite{lin2008}, the ratio $(\delta B_I + B_0)/B_0$ can be around 2-3 and,
correspondingly, $\delta B_I/B_0 \approx$ 1-2. Presumably, the ratio $(\delta
B_I + B_0)/B_0$ should not vary significantly as a CME moves in the corona and
heliosphere away from the Sun. Therefore, we obtain $\delta B_I \approx$ (1-2)
$B_0 \approx$ (1-2) G for $B_0 \approx$ 1 G in the corona.

Taking density $N \approx 3\times 10^6$ cm$^{-3}$ as an estimate for the
streamer belt at $R = 2$ R$_\odot$, and given $\delta B_I \approx$ (1-2) G, we
may estimate the drift velocity $V_d \approx$ (5-10)$\times 10^4$ km/s at the
CME boundary. At the same time, the thermal electron velocity $V_{Te} \approx
5\times 10^3$ km/s, for the coronal temperature $T_e \approx 1.5\times 10^6$ K,
i.e. the condition $V_d > V_{Te}$ is satisfied.

Thus, the type II radio bursts observed in the CME 2 and CME 3 events may be
caused by a drift current instability evolving at the CME frontal structure
boundary.


%
%
%
%
%
%

%
%
%
%

\begin{acknowledgments}


SOHO is a project of international cooperation between ESA and NASA. The
STEREO/SECCHI data are produced by a consortium of NRL (USA), LMSAL (USA), NASA/GSFC (USA),
RAL (UK), UBHAM (UK), MPS (Germany), CSL (Belgium), IOTA (France), and IAS (France). The Mark 4 data are courtesy of
the High Altitude Observatory/NCAR.
We thank V. Zaitsev for useful discussions.
The work was supported by the Russian Foundation of Basic Research (grants no. 09-02-00165 and 10-02-00607).

\end{acknowledgments}

%
%
%
%
%
%
%
%
%
%




%
%

\end{article}




\newpage

\begin{figure}
\caption{A and C,
slow CME 5 May 1997; B and D, fast CME 20 September 1997. A and B,
difference brightness isolines, $³L$ is the position angle, the
coordinate axes are in units of $R_\odot$. C and D, difference
brightness distributions depending on distance $r$, measured from
the CME frontal structure center (point O), along two different
sections 1 and 2, whose direction is shown by dashed lines in the
top panels. LASCO C2 data.}
\end{figure}

\begin{figure}
\caption{Velocities $u = V - V_{SW}$ relative to the surrounding
SW depending on the distance from the Sun center for the CME
frontal structure (light symbols) or for a shock ahead of the CME
(black symbols) in the propagation direction. Black triangles
inside the circles correspond to the shock front at, presumably,
the forming stage. Dashed curve is velocity $V_{SW}$ of
quasi-stationary slow solar wind in the streamer belt in
\cite{wang2000}. Solid curve is the Alfv\'{e}n velocity in the
streamer belt in \cite{mann1999}.}
\end{figure}

\begin{figure}
\caption{Difference polarization brightness profiles $\Delta P(R)$
at consecutive instants based on Mark 4 data for the 25 March 2008
CME. Light circle profiles correspond to the undisturbed corona.
To the right of each instant are shown the parameters indicating
the profile position in the image: its reference point ($PA_C$,
$R_C$) and direction (angle $\alpha$).}
\end{figure}

\begin{figure}
\caption{Difference brightness profiles $\Delta P(r)$ in a
coordinate system tied to the frontal structure at consecutive
instants based on COR1a data: A) 31 December 2007 CME; B) 3 June
2007 CME. The inscriptions are as in Figure~3.}
\end{figure}

%
%
%
%
%
%


\end{document}